\documentclass[12pt,preprint]{aastex}
\begin{document}

\title{X-ray and Optical Variations in the Classical Be Star $\gamma$ 
Cassiopeia: The Discovery of a Possible Magnetic Dynamo }

\author{Richard D. Robinson}
\affil{Catholic University of America \& Computer Sciences Corporation}
\affil{3400 N. Charles St., Baltimore, MD 21218}
\email{robinson@pha.jhu.edu} 

\author{Myron A. Smith}
\affil{Catholic University of America \& Computer Sciences Corporation/STScI}
\affil{ 3700 San Martin Drive, Baltimore, MD 21218}
\email{msmith@stsci.edu}

\author{Gregory W. Henry}
\affil{Center of Excellence in Information Systems, Tennessee State University}
\affil{330 10th Avenue North, Nashville, TN 37203}
\email{henry@schwab.tsuniv.edu}

\begin{abstract}
 $\gamma$\,Cas is a classical B0.5e star known to be a unique X-ray source by 
virtue of its moderate L$_x$ (10$^{33}$ erg~s$^{-1}$), hard X-ray spectrum, 
and light curve punctuated by ubiquitous flares and slow undulations. The 
peculiarities of this star have led to a controversy concerning the origin of 
these emissions; whether they are from wind infall onto a putative degenerate 
companion, as in the case of normal Be/X-ray binaries, or from the 
Be star itself. Recently, much progress has been made to 
resolve this question: (1) the discovery  that $\gamma$\,Cas is a
moderately eccentric binary system (P = 203.59~d) with unknown secondary type, 
(2) the addition of {\it RXTE} observations at 6 epochs in 2000, adding to 3 
others in 1996-8, (3) the collation of robotic telescope ({\it APT}) $B$ and 
$V$-band photometric observations over 4 seasons which show a 3\%, cyclical 
flux variation with cycle lengths of 55 - 93 days. 

We find that X-ray fluxes 
at all 9 epochs show random variations with orbital phase, thereby 
contradicting the binary accretion model, which predicts a substantial 
modulation.  However, these fluxes correlate well with the cyclical optical 
variations. In particular, the 6 flux measurements in 2000, which vary by 
a factor of 3, closely track the interpolated optical variations between 
the 2000 and 2001 observing seasons. 

The energy associated with the optical variations greatly exceeds the energy 
in the X-rays, so that the optical variability can not simply be due to 
reprocessing of X-ray flux. However, the strong correlation between the two
suggests that they are driven by a common mechanism.  We propose that this 
mechanism is a cyclical magnetic dynamo excited by a Balbus-Hawley 
instability located within the inner part of the circumstellar disk. According
to our model, 
variations in the field strength directly produce the changes in the
magnetically related X-ray activity. Turbulence associated with the dynamo
results in changes to the density (and therefore the emission measure) 
distribution within the disk and creates the observed optical variations.   

\end{abstract}


\section{Introduction}

Since its discovery in 1867, $\gamma$\,Cas has become the prototype of the
``classical'' Be stars. However, while the optical properties are
representative of the class, its behavior in the X-ray regime is not just
unusual but so far unique. Its mean L$_{x}$ is  a few times greater 
than any other Be or B-normal stars but at least a factor of 20 lower than 
in Be X-ray binaries. The X-ray light curve is especially interesting
in that it is composed of numerous, short-lived bursts 
(with durations of 10s to 1 minute) superimposed on a background 
(``basal flux") which 
varies in intensity by a factor of up to 3 over timescales ranging 
from hours to months. The X-ray spectrum 
is hard, being consistent with a thermal 10.5\,keV thin plasma, and it shows 
Fe lines at 6.7 and 6.9 keV.

In the past the commonly accepted explanation for the X-ray emission 
involved mass accretion onto a putative degenerate companion, either a
neutron star (White et al. 1982) or a white dwarf (Murakami et al. 1986).
There are some problems with this interpretation (see Smith, 
Robinson \& Corbet, [1998; SRC98] and
Robinson \& Smith [2000, RS00] for discussions of the binary hypothesis), and
Smith (1995) published an alternative picture in which the anomalous X-rays
were generated by energetic flares near the star's surface. To test this 
hypothesis SRC98 organized a coordinated observing campaign in 1996 March 
involving the {\it Rossi X-ray Timing Explorer (RXTE)} and the 
{\it Hubble Space Telescope}, where the 
Goddard High Resolution Spectrograph ({\it GHRS}), with large aperture,
was used to obtain high time resolution spectra covering 
40 \AA\ centered on the Si\,IV lines near 1400 \AA.~ The object was to look 
for correlations between X-ray variations and changes in the UV 
continuum and the Si\,IV line profiles. Since any UV variations are likely
to originate at or near the Be star, a strong correlation with the X-rays 
would provide evidence that the X-rays are also emitted from near the star.

The program proved to be highly successful and showed a pronounced {\em 
anti-correlation} between the intensity of the basal X-ray flux and the 
value of UV continuum near 1400 \AA\ (see SRC98), an 
anti-correlation with the Si\,III and Si\,IV (low-excitation) ions line 
strengths, and a direct correlation with the high excitation Fe\,V line 
strengths (Cranmer, Smith \& Robinson 2000).

The basal X-ray flux from the 1996 March observations was found to vary on a
$\sim$10 hour timescale (Fig.~\ref{lc9698}a). UV observations obtained during
{\it International UV Explorer (IUE)} campaigns in 1982 and early 1996,  
combined with the {\it GHRS} results, showed
a UV-continuum modulation of 27 hours. This is in agreement with the
expected rotation period of 22-30 hours given the star's known 
Vsin\,i and estimated radius (see, e.g., Cramner, Smith \& Robinson 2000). 
Thus, SRC98 suggested that the X-ray
variations are attributable to a rotational modulation of 
active magnetic complexes on the stellar surface.
To confirm this interpretation the {\it RXTE} was used on 1998 Nov 24-26 to 
search for a reoccurrence of the variation pattern over two rotation periods. 
A summary of part of this time history is given in Figure~\ref{lc9698}b. 
Comparison of the two panels of this figure
discloses that the light curve characteristics changed markedly 
between 1996 and 1998. The 1998 fluxes were substantially lower
than those seen in 1996 March, while the timescale of variation had decreased
to 3-4 hours. No clear signal for rotational modulation was discovered (see 
RS00). However, RS00 found that the X-ray fluxes did undergo a 
partial cessation every $\approx$7.5 hours. This behavior was intriguing 
since archival IUE data showed marked strengthening of the high-velocity
Si\,IV and C\,IV wind features (DACs) with the same cyclicity. 
The rapid timescale for the basal flux variability also points to a rapid 
evolution of the activity centers, which would mask any rotational modulation.

The dramatic changes in these X-ray light curves were unexpected. The 
authors concluded that more observations were needed in order to quantify 
the timescale and form of these variations. It was also hoped that the nature
of the variations could be better understood by comparing the X-ray
results with other parameters of the star, such as the one-armed density 
enhancement in the Be disk, which is known to be responsible for the
 5 - 7 year 
cyclic variations of the H$\alpha$ emission profile (Telting \& Kaper 1994, 
Berio et al. 1999). Thus, an observing  program using the Automated 
Photometric Telescope ({\it APT}) was initiated to monitor the optical 
variability of $\gamma$\,Cas in the Johnson  $B$ and $V$ bands. New 
{\it RXTE} observations were also carried out in 2000, consisting 
of 6 time sequences spaced at increasing intervals from 1 week to 5 months.  
It is these optical and X-ray data that will be examined in this paper.

In $\S$\ref{obs} we describe briefly the X-ray and optical observations and 
reduction procedures. The long-term variations are then described in 
$\S$\ref{ltvar}. Here it is shown that both the X-rays and optical
fluxes show cyclic fluctuations on a timescale of 55 - 93 days.
The X-ray and optical variations also appear to be strongly correlated. 
The short-term variations are then discussed in $\S$\ref{stvar}. Here we 
examine the short-term changes in the X-ray flux and compare them with similar
variations seen in the optical region. Many of the X-ray time sequences appear 
to have broad flux minima which are repeatable. These features are used as 
markers in attempting to determine the stellar rotation rate. The time 
sequences are also searched for the cyclic decreases in X-ray flux which
were found in previous data sets. In $\S$\ref{disc} we discuss the
consequences of the observations.

\section{Observations and Data Reduction} 
\label{obs}

\subsection{X-ray Data}

The X-ray data were obtained with the {\it RXTE} satellite using the 
Proportional 
Counter Array (PCA), which detects photons in the 2-30 keV range. A summary
of the observations is presented in Table\,1. A total of 6 time sequences 
were obtained, each approximately 27 hours in duration, which is near the 
expected rotation period of the star. Since a primary goal of the 
program was to determine the timescale of the long-term X-ray variations, 
the visits were spaced at progressively increasing time intervals, ranging 
from 1 week to 5 months.

The PCA data were reduced using standard procedures within the FTOOLS reduction
package, as described on the {\it RXTE} project {\it Cookbook} 
website\footnote{heasarc.gsfc.nasa.gov/docs/xte/recipes/cook\_book.html}. 
The PCA has been slowly aging over the years. 
In an effort to decelerate this process the project periodically
rests detectors by turning them off. During our observations various
combinations of the PCA proportional counter units (``PCUs"), typically
three, were operating at any one time and this reduced the effective total
count rate from our target correspondingly. More importantly, the energy
dependence of the background models for each of the detectors had diverged.
In fact, on 2000 May 13, a month before our fifth visit, the propane layer 
peeled away from the PCU0 unit. This event increased the soft background
level for that unit. A revised background model based on new calibrations 
of this unit was not available at the time of our reductions. Therefore, we 
reduced the net fluxes of our observations for each PCU separately,
and the color information from PCU0 was ignored in our last two data sets.
After the initial reductions, the source count rate was adjusted
to mimic that expected from a full 5-unit PCA, so that the values could be 
compared with fluxes obtained during the 1996 and 1998 observing programs.

\subsection{Optical Observations}

  Our photometric observations of $\gamma$\,Cas were obtained over four 
observing
seasons between 1997 September and 2001 February with the T3 0.4~m Automatic
Photometric Telescope ({\it APT}) at Fairborn Observatory in the Patagonia 
Mountains of southern Arizona.  This APT uses a temperature-stabilized EMI 
9924B
bi-alkali photomultiplier tube to acquire data through Johnson $B$ and $V$
filters.  The APT is programmed to measure stars in the following sequence,
termed a group observation:  {\it K, sky, C, V, C, V, C, V, C, sky, K}, where
$K$ is a check star, $C$ is the comparison star, and $V$ is the program star.
We used HD 6210 = HR 297 ($V$ = 5.80, $B-V$ = 0.58, F6\,V) as the comparison
star and HD 5395 = HR 265 ($V$ = 4.63, $B-V$ = 0.96, G8\,IIIb) as the check
star for our observations of $\gamma$\,Cas.  Typically two or three group
observations were acquired each clear night at intervals of two to three 
hours. 
In the fourth observing season (2000-1), up to several dozen group
observations were obtained on four nights of more extensive monitoring.  To
keep coincidence count corrections small for these bright targets, a roughly
3.8 magnitude neutral density filter was used for all integrations during the 
second through the fourth observing seasons. Integration times were 10 seconds 
for the check star and $\gamma$\,Cas and 20 seconds for the comparison star and 
sky measurements.  During the first observing season (1997-98), a 4.8 magnitude
neutral-density filter was used for $\gamma$\,Cas and a 1.2 magnitude filter 
for everything else.  This complicated the reduction of the observations to
differential magnitudes for the first season and led to a small
zero-point offset relative to subsequent seasons (see below).

  Three variable minus comparison and two check minus comparison differential
magnitudes in each photometric band were computed for each group observation
then averaged to create group-mean differential magnitudes.  The group means
were corrected for differential extinction with nightly extinction
coefficients, transformed to the Johnson system with yearly mean
transformation coefficients, and treated as single observations thereafter.
The external precision of the group means, based on standard deviations for
pairs of constant stars, is typically $\sim$0.004~mag on good nights with this
telescope.  Group mean differential magnitudes with internal standard
deviations greater than 0.01~mag were discarded to filter observations
taken in non-photometric conditions.  A total of 921 $B$ and 927 $V$ group
mean differential magnitudes for $\gamma$\,Cas were obtained over the four
observing seasons\footnote{The individual Johnson BV photometric observations 
are available at http://schwab.tsuniv.edu/t3/gammacas/gammacas.html   
and will be eventually published in full. } 
Further details of telescope operations and data-reduction
procedures can be found in Henry (1995a,b).

\section{Long-Term Variations} 
\label{ltvar}

\subsection{X-rays}
\label{xltvar}

The representative X-ray intensity at a given epoch is assumed to be 
characterized by the
mean value of the X-ray flux evaluated over the 27 hour rotation
period of the star. This mean value was derived for each of the 6
time sequences in the present study as well as the two observations taken
in 1996 and one taken in 1998. The results are presented in Table\,1, 
which shows that the X-rays can vary by a factor of up to 3 in intensity. 

A periodogram analysis was carried out on the 9 sample dates and solutions
were found at 70.1 days and 84.4 days (see Fig.~\ref{ltxvar}). In both 
cases the light curve was nearly sinusoidal. While these plots support
the view that the X-ray fluxes have a long-term periodicity,
the number of samples is insufficient to make a compelling case that 
the X-ray variations are strictly periodic. In fact, in $\S$\ref{xoptcomp}, 
we will introduce optical evidence that these X-ray variations 
are cyclic with a still unknown mean timescale.

\subsection{Optical}
\label{optltvar}

The time histories of the $B$ and $V$ intensities for all 4 APT 
observing seasons
are shown in Figure~\ref{optsum}, while a summary of the characteristics 
of these data are given in Table\,2. In all cases these time histories 
show a pronounced sinusoidal variation. A periodogram analysis for each
observing season shows a variable period, 
decreasing from 61 days for the 1997-1998 season to 54 days
in the 1998-1999 season and then increasing thereafter, reaching 93 days
for the 2000-2001 season (see Table\,2). 

In addition to the cyclic 55 - 93 day periodicities, there also appears to be
significant changes in the average magnitude from one observing season 
to the next.
Since the first season was obtained through different neutral density filters
than the following observations, it is not possible to compare the optical
fluxes of those data with later seasons. However, data for the last three 
seasons were obtained with identical instrumental configurations, so these 
long-term variations are probably real. 
Such variations are in fact typical of early-type ``classical" Be stars
(Pavlovski et al. 1997, Moujtahid et al., 2000). Although the reason(s) for the 
variation are unknown, they are generally thought to be associated with an 
evolution of the disk structure (Hirata 1983).  

In Figure~\ref{bvcomp} we examine the relation between the $B$ and $V$ fluxes 
during the four observing seasons. To facilitate the comparison we have 
averaged all observations taken during an individual night and converted the 
fluxes to percentage deviation from the average intensity measured during 
that observing season. The dashed line in each plot represents the case where
the change in $B$ equals the change in $V$, so there 
would be no color change  associated with the change in intensity. The solid 
lines represent a linear least squares fit of $\Delta B$ against 
$\Delta V$. The slope of 
this fit is indicated on the plot. In all four observing seasons the slope is 
less than 1, indicating a greater fluctuation in $V$ than in $B$. This result 
is consistent with other observed color variations. For example, 
Horaguchi et al. (1994) show a strong positive correlation between the (B-V) 
color and the $V$ band intensity of $\gamma$ Cas, in which (B-V) increased by 
15\% during a 45\% increase in the $V$ band flux between 1960 and 1989. 

To see whether the small slopes presented in Figure~\ref{bvcomp} are 
statistically significant, and not caused 
by the large uncertainties in the measured values of $\Delta~B$ and 
$\Delta~V$, we repeated 
the analysis using 2 and 3-day averages. The results were nearly identical to 
those found for the one-day averages. We also performed a statistical test 
in which we calculated the probability of obtaining the measured slopes from 
data in which the slope was actually equal to 1. To do this we used a 
random number generator to create a set of 100 $\Delta B$ and 
$\Delta V$ samples with values ranging from -1.5\% to 1.5\% and 
for which $\Delta B= \Delta V$. A second random
number generator was used to add normally distributed uncertainties to the 
data. The distribution of uncertainties was centered on 0 and had a 
$\sigma$ of 0.3\%. A least squares 
fit was then performed on the synthetic data set and the slope was tabulated.
Repeating the process 20,000 times, we were able to empirically establish the 
probability of obtaining any given slope. The probability distribution was 
gaussian in shape and was centered at a slope of 0.9 with a FWHM of 0.1. The
offset of the centroid slope from the expected value of 1.0 was caused by the 
large uncertainties in the values of the independent variable ($\Delta V$). 
When these 
uncertainties were reduced in the simulation, the center of the probability 
distribution shifted to 1.0, as expected. From these simulations we find that
the probability of measuring a slope of less than 0.8 is only 2.5\%, while 
the probability of a slope less that 0.75 is only 0.07\%. Thus, while the
distribution obtained during the first season (Fig.~\ref{bvcomp}a) is 
compatible with $\Delta B = \Delta V$, those obtained in other seasons
are consistent with $\Delta B < \Delta V$.

To estimate the uncertainties in the  slopes, 
we determined the rms deviation of the measured points from regression lines 
having slopes ranging from 0.3 to 1.5 and pivoting around the point
$\Delta B = \Delta V = 0$. The derived distribution was a rather
flat parabola centered at the slope deduced from the least squares fit. The
range of acceptable slopes was taken to be those values where the rms deviation
was less than 1.5 times the minimum rms value. This range is indicated in
Fig.~\ref{bvcomp}. 

Future work by one of us (GWH) will test these results further by extending 
the photometric monitoring into the near-IR.

\subsection{Comparing Optical and X-ray Variations}
\label{xoptcomp}

The sinusoidal shape and 75 day period found in the optical data during 
the 1999-2000 observing
season is remarkably similar to the 70 day variations found for the X-ray
fluxes (Fig.~\ref{ltxvar}). An obvious question, therefore, is whether 
the variations in the
two wavelength regimes are correlated. Unfortunately, it is not possible to 
compare the two data sets directly, since only 5 of the 9 X-ray time sequences
have simultaneous optical coverage and most of those were obtained during the
end of the optical observing season, where the uncertainties in the optical
magnitudes were large. 

An alternative approach is to determine an empirical model for the optical
variations during the last two observing seasons and then compare  
the X-ray fluxes with the prediction of this model. 
We know from the period analysis discussed in section $\S$\ref{optltvar}
that the  1999/2000
season has a period of roughly 75 days, while the 2000/2001 period was about
93 days. The optical variations were therefore modeled as a sine
wave with a period linearly increasing with time. We also assumed
a phase and mean intensity which varied linearly with time, since the overall
flux increased between the two seasons. Thus, the assumed light
curve had the form:

     m$_V$ = C$_1$(t) + D~sin( C$_2$(t) + 2$\pi$\,t/C$_3$(t) )
     
\noindent
where:   C$_i$(t) = A$_i$ + B$_i$t ~ ; ~ (i=1,2,3) \\

Here A$_i$, B$_i$ and D are all constants which were manually adjusted 
to fit the
observations and t is the time (in days) from a reference date (HJD=2,451440). 
To increase the S/N during the fitting process  the average $B$ 
magnitudes were 
scaled to match the $V$ magnitudes and the data were averaged into 3 day bins. 
The resulting fit is shown in Fig.~\ref{optx} and is remarkably good 
considering the crudeness of the model. The parameters of the fit were:

\noindent
\hspace*{1in} C$_2$(t) = phase = 0.85$\pi$ - 0.0015$\pi$t \\
\hspace*{1in} C$_3$(t) = period = 65 + 0.027t \ \ \ days \\

\noindent      
Note that the period is 65 days at the start of the 1999/2000 season and
increases to about 79 days at the end of the 2000/2001 season. This is somewhat
shorter than was obtained when fitting the individual seasons with a sine 
curve. The difference is probably due to the relatively large change
in phase during an observing season.

  We next computed a numerical model for the X-ray light curve by using the 
values of period and phase which were found for the optical data and adjusting
the average intensity (C$_1$) and the amplitude (D) to fit the observed X-ray 
fluxes. The results, shown in Figure~\ref{optx}b, exhibit a remarkably strong
correlation between X-ray and optical fluxes. We note parenthetically that
the year-to-year increase in the optical data is not reflected in the X-ray 
fluxes. This is consistent with the fact that such variations are also 
commonplace in Be stars without strong X-ray emission (Pavlovski et al., 1997).

\section{Short-term Variations }
\label{stvar}

\subsection{General Properties of the X-Ray Flux}
\label{stxvar}

In Figure~\ref{xsum} we show the X-ray variations for each of the 6 
time sequences taken during the 2000 program. 
As discussed by SRC98 and RS00, the X-ray flux is composed of two 
components. The first consists of numerous short duration 
bursts, termed shots, with lifetimes of $\sim$10s to several minutes. 
These are superimposed on a background, or basal emission, which varies on 
timescales of 30 minutes to $\sim$ 10 hours. The basal emission can be
thought of as the minimum flux at any given time. It was determined 
by two independent techniques (see RS00). Briefly, the first of these 
techniques relies upon determining the median flux during time periods 
(of several minutes duration) when no significant shot events occur and 
then linearly interpolating between these samples. The second technique 
attempts to remove the shot events from the time sequence and then takes 
the average residual flux as the basal component. 
Both techniques produce similar
results, although the second generally leads to a greater variability since
groups of shots often merge into apparently longer-term variations. 

  Our estimates of the basal fluxes are presented in Figure~\ref{xsum}. 
Comparing the different time sequences shows that the character and timescale
of the variations can change substantially from one visit to the next. 
However, some trends can also be detected within the data. 
For example, visits 1, 2, 3 and 5 are all taken during relatively low-flux 
epochs, when the average count rate was 60 counts~s$^{-1}$ or less (see 
Table\,1 and Figure~\ref{optx}). In all cases these time sequences show a
wide range of basal fluxes and a rather long timescale for the variations,
similar to the behavior seen during the 1996 March observations (SRC98). 
In every case there was at some point a long-duration flux minimum (marked 
by a vertical line in Figure~\ref{xsum}), which will be discussed in more 
detail below. 
During visit 2, when the mean flux was low, there were times when the 
fluxes dipped to below 15 cts~s$^{-1}$, and the instantaneous X-ray output 
approached the normal levels for a early-type Be star. On one occasion the
hardness ratio (formed from fluxes summed over 7.6 - 12 keV to 
the integrated 2 - 4.1 keV flux) also 
dropped significantly over several minutes to a new stable value. In
contrast, the time sequences obtained during visits 4 and 6 were obtained 
during flux maxima, when the average {\it RXTE} fluxes were nearly 100 
counts~s$^{-1}$. The basal fluxes show much less variation for these 
epochs, and there is no indication of the broad flux minima. 

Variations of the basal flux occur in all of the light curves and last 
from about 30 minutes to several hours. 
Details of several representative time sequences are shown 
in Figure~\ref{xshort}. 
The timescale for these variations are much too small for them to be caused
by rotational modulation, and their presence emphasizes the extremely dynamic
nature of the X-ray source region. Short-duration events appear to be caused 
by dramatic increases in the production rate of individual shots. The longer 
term events often contain periods of several minutes or longer during which 
the shots temporarily disappear. The enhanced flux levels during this time 
therefore reflects a rapid evolution of the basal source region itself.

  An inspection of the hardness ratios shows to first order that the 
hardness typically does not vary, even when the integrated net fluxes change 
by large factors. There are exceptions 
to this, and there is a statistically significant tendency for the 
hardness ratios to increase slightly when the net fluxes increase for several 
hours. There is also a striking case during visit 4 where the hardnesses 
decreased from an abnormally high value to normal in 20 minutes.
These results are in accord with our analysis of the 1996 and 1998 data sets 
(see SRC98, RS00).

\subsection{Redetermining the Rotation Period} 
\label{rot}

The broad flux minima seen in visits 1, 2, 3 and 5 are very similar to a 
feature seen during the 1996 March observations (see SRC98). The duration for 
this feature is about 7 hours and may therefore be caused by the rotational 
modulation of a structure on or near the surface of the star, reminiscent 
of a solar coronal hole. The shape of the time series for this feature 
appears to be stable for moderately long periods of time, as shown in 
Figure~\ref{mincomp}. In this plot we compare 30 minute averages of the 
time sequences obtained during 
visits 1 and 3 and assume a 27 hour rotation period. The light curve for 
visit 3 has been shifted so that the flux minimum matches the minimum seen 
during visit 1. Comparisons between other time sequences which have 
the feature 
show a similar agreement. The stability of the feature suggests that it is 
formed in a relatively long-lived structure (the feature may simply be masked 
by enhanced activity from other longitudes on the star
during the two visits where it was not apparent) and can 
therefore be used as a marker from which a rotational period could be deduced. 
Such markers were used by SRC98, who used the feature in the 1996 March 
{\it RXTE} 
data together with {\it ASCA} satellite data taken 11 days earlier to determine
a rotation rate of 1.125 days (27 hours). This is very similar to the 
period of 1.12277 days derived from UV variations (RS00).

The relative times at the center of the flux minimum for each of the 
relevant time sequences are given in Table\,1; these are referred to the 
first sequence in 2000 January, However, a simple period analysis shows 
that there is no unique rotation period which can account for all of these 
time intervals. This implies that the feature is either dissipating and 
reforming at 
new longitudes or that it is drifting in longitude ($\phi$)  at a rate which 
changes with time, so that the longitude at any given moment is given by:

 $\phi$(t) = $\phi$(0) + Ct$^{\beta}$ \\

\noindent
If $\beta$ were equal to 1, so that the drift rate were a constant,
then the markers would still be periodic except that the measured period would 
differ from a true rotation period. Using the measured time 
intervals from Table\,1 we find that a suitable set of parameters is given by:

\noindent
\hspace*{1in} apparent rotation period at t$_{o}$ = 25.18 hours \\
\hspace*{1in} C = 0.241 degrees hour$^{-1}$ \\
\hspace*{1in} $\beta$ = 1.2 \\

\noindent We stress that these figures refer only to this set of
X-ray data and should
not necessarily to be preferred over those fits to earlier data, since we 
expect that the values of C and $\beta$ would vary with time.

\subsection{A Search for Periodic Decreases in the X-ray Flux}

\label{flxdec}

In their analysis of the 54-hour X-ray time sequence obtained in 1998 Nov, RS00
discovered the presence of cyclic decreases (``lulls") in the X-ray flux which 
occurred every $\approx$7.5 hours. These decreases were also found  
in the 1996 March observations. To check for a persistence of the X-ray 
lull pattern, we carried out the same analysis on the six data strings 
obtained in 2000. Surprisingly, with the possible exception of time sequence 
5, no evidence  for lulls with this period were found the data,  as shown in 
Figure~\ref{invccor}. In Fig~\ref{invccor}a the autocorrelation analysis 
of the 1998 Nov sequence is shown for reference. Light curves 
obtained during low flux epochs (visits 1 - 3) showed no indication of 
periodic lulls. However, some evidence 
of cyclicity was seen during the high-flux visits, but on a much shorter 
timescale, i.e., 3.5 hours for visit 4 (Fig~\ref{invccor}b) and 5.8 hours 
for visit 6 (Fig~\ref{invccor}d).  The general indication 
is that the lull-cycles can develop, last for several months or more,
and then disappear. Lull-cycles from these different ``episodes" apparently
do not have to take on a special value (like 7-hours) as had been tentatively 
concluded in RS00. We are as yet no closer to a fundamental understanding 
of this peculiar phenomenon.

\subsection{Short-Term Variations in the Optical Flux}
\label{optext}

  Most of the {\it APT} observations consisted of only 1 or 2 samples per night.
However, on 4 nights during the 2000/2001 observing season more intensive
observations were carried out, covering 4 hours when the source was near the 
zenith. The resulting light curves are shown in Figure~\ref{shortopt}.
Despite the large scatter, it is apparent that there are real 
variations on the 1 - 3\% level which have timescales of several hours
(see especially Figure~\ref{shortopt}c). 
X-ray variations can also occur on these timescales (e.g. 
Fig.~\ref{lc9698}b). Unfortunately, none of the extended optical data sets 
were taken precisely during an {\it RXTE} visit, 
so the short-term X-ray/optical association cannot be examined in detail.

\subsection{ H$\alpha$ Variability }

   During 1928 - 34 and from 1969 to the present the H$\alpha$ profile of 
$\gamma$\,Cas has been dominated by cyclic variations of the V/R emission
ratio (e.g., Doazan et al. 1983). These variations are now widely believed
to be due to a dynamical instability in the disk which excites a one-armed
density wave which precesses around the star on a timescale of 5 - 7 years.
This feature has been imaged interferometrically in H$\alpha$ light and has
been shown to rotate around the star on this timescale (Berio et al. 1999).
During the year 2000 the V/R ratio decreased from a value of about 1.3 to 
1.0 (Peters 2001). During this time the equivalent width of the line also 
decreased by about 3\%, continuing a slow downward trend since its gradual 
ascent to maximum in 1994 (Pollmann 2001).

  The original impetus of our 2000 {\it RXTE} program was to measure a 
timescale for X-ray changes in order to determine whether the epochal 
X-ray flux levels were related to the passage of the one-armed circumstellar 
feature. This might occur if, for example, the density of the arm affected 
an interaction between a putative stellar magnetic field and a 
magnetic field entrained or 
self-generated in the disk. We can now say (cf. Fig.~\ref{optx}) that there 
is no evidence for a year-to-year trend in the X-ray flux which could be 
ascribed to a disk-arm precession cycle.

The discovery of the cyclic variability in the optical continuum leads to the
question of whether this cycle is also present in the H$\alpha$ data. 
To help us address this question,  K. Bjorkman 
and A. Miroshnichenko kindly put at our disposal a series of 88 H$\alpha$ 
line profile observations obtained at the Ritter Observatory during an 18 
month period from late 1999 to early 2001. Measurement of these data show 
that there are no apparent cyclic variations in the H$\alpha$ equivalent 
widths (EW) on any timescale, including one near 70 days. The same holds for 
the shape of the normalized profile, which shows only a steady, long term 
decrease in the amplitude of the violet peak. However, we note that 
equivalent widths are defined with reference to the continuum flux. Therefore,
the H$\alpha$ {\em fluxes} should actually have a variation which is similar
to that of the optical continuum, provided cyclic EW variations were 
not lost in the noise.
To test whether 3\% EW variations were actually detectable in these data
we added synthetic  sine curves with a 3\% amplitude and then analyzed them 
with a periodogram tool, PDM, in {\it IRAF}. The calculation was then 
repeated with the sine curve starting at different phases. For each 
of these trials we were able to detect the input signal at about the same, 
2$\sigma$, level. This implies that an {\it absence} of a modulation in the 
H$\alpha$ fluxes, seen as a reflex in the equivalent width data, can be 
verified. This is a nontrivial point. Since the disk is certainly optically 
thick in H$\alpha$ and thin in the continuum, the two could conceivably show 
different responses.

\section{Discussion} 
\label{disc}

\subsection{Examining the Binary Accretion Hypothesis}

As mentioned in the introduction, it is commonly suggested that the X-rays
from $\gamma$\,Cas originate from mass infall onto a degenerate companion.
For years this hypothesis could not be directly tested because no evidence for 
a companion existed. However, recently Harmanec et al. (2000) have reported, 
on the basis of periodic velocity variations in the H$\alpha$ and He\,I lines,
that $\gamma$\,Cas is likely to be a binary. Their solution for the velocity
curve shows that it is in a moderately eccentric orbit with a period of 
203.59 days and its companion has a mass $\sim$0.5-2 M$_{\odot}$. The
component stars in such a system would be separated by about 0.8 AU.

Standard Bondi-Hoyle accretion theory suggests that the X-ray luminosity from 
an accreting degenerate
companion will vary as $\rho V_{rel}^{-3}$, where $\rho$ is the local density
and $V_{rel}$ is the relative velocity between the star and the accreting 
gas. Since these properties vary with orbital phase, comparing the observed
{\it RXTE} X-ray flux with expected variations based on the orbital ephemeris
provides a clear test of the degenerate companion model. This comparison is 
done in Fig.\ref{bincomp}. Here the expected X-ray flux was calculated  
using the formalism for wind accretion onto a degenerate star described by 
Waters et al. (1989), in conjunction with the  Harmanec et al. orbital 
ephemeris with e=0.26. In this calculation it is assumed that the density 
varies as r$^{-2}$ and that the wind has reached its terminal velocity 
($\sim 1800$ km s$^{-1}$), so that $V_{rel}$ does not vary with orbital phase. 
Even stronger phase variations are expected if the companion is embedded
in the stellar Keplarian disk, since the density there varies as r$^{-n}$,
with n=3.3 to 4.5 (Waters et al., 1991).  In clear contrast to the results 
presented in Fig.~\ref{ltxvar}, Fig.\ref{bincomp} shows that the 1996 - 2000 
{\it RXTE} observations have a random scatter when folded over the Harmanec 
binary period. We believe that this result argues strongly against the idea 
that the X-rays originate from the companion. 

Recently, Mironishenko and Bjorkman (2002, private comm.) have used new data to
confirm the Harmanec et al. orbital period, but suggest that the ellipticity 
may be small. If this turns out to be the case, then the  X-ray variations 
predicted in Fig~\ref{bincomp} would disappear. In this scenario, variable 
X-ray emission from a companion would require changes in the wind mass loss 
rate or velocity. While this is conceivable, it it is unclear how or where 
these wind changes would originate.
If the companion were embedded in the disk, then the orbit is likely
to have at least a small inclination to the plane of the disk (e.g. Waters 
et al., 1989). Since Be disks are thin, the star would then be subject to 
substantial phase-related density and relative velocity variations, with 
consequent phase related X-ray production, even for a circular orbit. 

The current study provides two additional pieces of evidence against 
the companion as the source of the X-rays. The first involves the simple 
fact that the period of both the optical and X-ray variations {\em changes 
with time}, contrary to the behavior expected from a stellar companion. 
The second concerns the strong correlation  between the X-ray and 
optical changes ($\S$\ref{xoptcomp}). Even if we assumed that the Harmanec 
et al. (2000) analysis were erroneous and that the star had a companion 
with a 70 day period, an X-ray source at that distance would be unable to 
substantially affect the Be star's optical flux through irradiation. In 
fact, known high-mass X-ray binary systems with periods near 70 days and 
X-ray fluxes which are 2 - 4 orders of magnitude larger than those seen in 
$\gamma$\,Cas are not associated with any correlated optical variations 
(Liu, van Paradijs, \& van den Heuvel 2000). The idea that the optical 
variations are caused by reprocessing of the X-rays photons is also 
contradicted by energy considerations, as shown in $\S$\ref{optsite}.

 Another possibility is that the X-ray flux variations from a mass-accreting 
degenerate companion are driven by density changes associated with the optical 
variability (e.g, changes in the inner disk structure - see 
$\S$\ref{optsite}). In this case, the changing periods and close correlation 
of fluxes could conceivably be explained provided the companion were embedded 
within the disk and the disturbance propagates rapidly from the site of the 
optical emission near the Be star to the companion.   Disturbances can 
propagate rapidly (at the Alfven velocity), but only along the magnetic 
field direction, which 
is primarily azimuthal within the disk (see $\S$\ref{dynamo}).  Radially 
directed disturbances will propagate near the local sound speed, $c_s$.
For example, in the dynamo discussed in $\S$\ref{dynamo} the relevant 
velocity is $\alpha c_{s}$ (Vishniac 2002), where $\alpha$ is the effective
viscosity  and is probably less than one for Be stars (Okazaki 2001).
Assuming a mean disk temperature of T=10$^4$ K, one finds 
$c_{s} \approx$ 13 km s$^{-1}$. Thus, for a system with a 
separation of 0.8 AU, there would be a phase lag of about 100 days 
between the optical and X-ray light curves.

\subsection{The site of optical variations}
\label{optsite}

  The strong correlation between the X-ray and optical variations suggests
a common origin. It is tempting to postulate that the optical 
variations arise from the reprocessing of the X-ray photons as they impinge
the stellar photosphere. However, simple energetic considerations shows that 
this is not tenable. The bolometric luminosity of $\gamma$\,Cas is about 
10$^{38}$ erg~s$^{-1}$ 
(e.g., Harmanec 2000), and the epochal average X-ray flux according to our 
{\it RXTE} data is in the range 0.4 - 1.1$\times$10$^{33}$ erg s$^{-1}$. 
Thus, a 3\% variation of the bolometric flux corresponds to an energy input
of more than 10$^3$ times the observed X-ray energy. Using a Kurucz theoretical
LTE spectrum of a star with an effective temperature of 27,000~K and
log\,g=4, we find that the flux contained in the $B$ band and longer 
wavelengths equals $\sim$4.3\% of the bolometric flux. Thus, even if the 3\% 
optical variations are restricted to the $B$ and $V$ bands, they still
involve energies which are 100 times those observed in the X-rays.

If the optical variations are not produced directly by the X-rays, then 
it is likely that the two are 
driven by the same engine. One possibility is that the optical variations
come from the surface of $\gamma$\,Cas. In SRC98 we proposed a model 
in which the X-rays were produced by electron beams directed toward the star. 
These beams impulsively heat the stellar photosphere (at densities of 
$10^{13} - 10^{14}$ cm$^{-3}$) to temperatures near 10$^8$ K and produce
the shots. The heated material expands rapidly into the overlying magnetic 
canopy, where it radiates slowly and is responsible for the basal flux
(see RS00 for more details). In previous studies we have assumed that all
of the beam energy goes into the X-ray emitting plasma. However, it is 
possible that a significant amount may also heat the stellar photosphere 
to more moderate temperatures. 
In addition, the physical processes which result in the electron beams
(e.g., Smith \& Robinson 1999) may also
produce high energy protons, conduction fronts and/or Alfven waves (e.g.,
Ulmschneider, Priest \& Rosner 1991), all of which could heat the stellar 
atmosphere and account for the optical variations. 

  One problem with this general scenario is that a heating of the stellar 
atmosphere necessarily results in a decrease in the $(B-V)$ color of the star.
This is contrary to the observations presented in $\S$\ref{optltvar}, which 
show an {\em increase} in 
$(B-V)$ with increasing luminosity during the $\sim$70 day cycle. A heating 
also implies that the optical variations should be correlated with greater 
variations in the ultraviolet. To check this, we have compared the 
UV continuum fluxes extracted from 14 available high-dispersion 
large-aperture, LWP-camera observations from the {\it IUE} data archive at
MAST.\footnote{Multi-Mission Archive at Space Telescope Science Institute,
in contract to NASA.} These fluxes show no credible periodic variations
in the near UV to 
$\pm{2}$\%. A similar exercise for 164 large-aperture SWP-camera observations 
shows no reliable variations of the rotation averaged far-UV intensity with 
amplitude greater than about 0.5 percent. Thus, it appears unlikely that the 
optical flux modulations originate from the stellar surface. 

An alternative possibility is that the optical variations arise within the 
circumstellar disk. This is a commonly invoked explanation for other long-term 
optical variations seen in Be stars (e.g. Hirata 1983). For example,
Horaguchi et al. (1994) document a tight correlation between enhanced
V band flux and increased $(B-V)$ color of $\gamma$\,Cas during the disk 
building phase which occurred between 1960 and 1990. The 
relationship between $B$ and $V$ reported by Horaguchi et al. is also 
similar to the results presented in $\S$\ref{optltvar}, i.e. 
$\Delta B$/$\Delta V$ = 0.76 $\pm{0.08}$ (Fig.~\ref{bvcomp}). To understand 
this behavior we note that the disk, which has a density-averaged temperature 
of about 10,000~K (Millar et al. 2000),  exhibits a
progressively increasing contribution to the observed flux with increasing 
wavelength. It contributes some 0.05\% at 1900 \AA~ (Stee 2001), a few 
percent near 4800 \AA,~ about 20\% near 6500 \AA~ (Stee \& Bittar, 2000, 
Horaguchi et al. 1994) and virtually 100\% above 2 $\mu$m (Waters et al. 
2000). Thus, variations in the disk brightness would 
naturally explain both the increase in $(B-V)$ color with increasing optical 
flux and the lack of substantial UV variations. However, it should be 
pointed out that the observed flux variations are probably not simply the 
result of mechanical heating through, for example, resistive dissipation of 
disk turbulence. Assuming a Keplerian disk of mass $\sim 10^{-9}$ M$_{\odot}$ 
(Waters et al. 2000), one finds that the total orbital kinetic energy of
the disk is only $\sim 10^{38}$ erg. Thus, the disk does not appear to 
have enough reserved energy to allow it to power changes in radiative 
flux, amounting to 10$^{35-36}$ erg s$^{-1}$, for more than several minutes.
Instead, we assert that the observed energy most likely comes from the 
stellar radiation 
field, with the physical processes in the disk simply modulating the 
amount of energy which is absorbed and released by changing the  
optical thickness of the disk. This idea will be discussed in more detail in 
the $\S$\ref{dynamo}. 

  In our picture the X-ray emission from $\gamma$\,Cas comes from energy
stored within unstable magnetic fields.
Long-term changes in the X-ray emission should therefore reflect long-term 
changes in the magnetic fields. This suggests in turn that the X-ray emission 
is being driven by a magnetic dynamo. The close 
agreement between the X-ray and optical variations implies that this dynamo 
is also the controlling mechanism for the optical variations.  A physical 
processes explaining how this might happen is discussed in the following 
section.

\subsection{A Possible Dynamo}
\label{dynamo}

  Smith and Robinson (1999) suggested that many of the observed X-ray and UV 
characteristics of $\gamma$\,Cas could be explained by the dynamical
interactions between putative magnetic fields on the star and its disk. 
While the short-term variations in the basal flux  and burst rate 
($\S$\ref{stxvar}) can be attributed to an inhomogeneous magnetic structure, 
the long-term variations ($\S$\ref{ltvar}) are most likely caused by cyclic
changes in the magnetic field strength and/or area coverage on the star or  
within the disk. We believe that this implies the presence of a 
magnetic dynamo, as mentioned above, and can think of no other explanation
that fits our observations. The site of this dynamo is uncertain. It may 
occur on the Be star itself. Such a dynamo would be completely different 
from that operating on late type stars such as the Sun, since a 
conventional $\alpha - \Omega$ dynamo requires convective motions, 
which are not present on $\gamma$\,Cas. It is possible that Coriolis forces on 
this rapidly rotating star may substitute for the convection, as suggested by 
Airapetian (2000). However, the fact remains that $\gamma$\,Cas is currently
the only known Be star which shows this type of X-ray and optical variations,
and there is nothing particularly unusual about the star itself other than
its not atypical Be character. 

The alternative is that the dynamo operates in the circumstellar disk, which is
among the densest of all known Be stars (Poeckert \& Marlborough 1978, 
Lamers \& Waters 1987, Telting 2000). As described in Balbus and Hawley (1998),
a disk dynamo is substantially different from a classical stellar dynamo. 
In a star, the convective turbulence that is responsible for amplification 
of an existing seed field (the so-called dynamo $\alpha$ effect, not to 
be confused with the Shakura-Sunyaev viscosity $\alpha$, both used in 
disk accretion studies) is a global stellar property. These motions contain
much more energy than the magnetic fields and are therefore unaffected by 
the growth in the field strength. 
In a disk, however, the turbulence is produced by the interaction 
of a seed field (presumably coming from the star) and the Keplerian shear 
within the disk through the magneto-rotational instability of Balbus \&
Hawley (1991), which is expected to operate to some extent whenever
a magnetic field is embedded in a Keplerian disk (Balbus \& Hawley 1998).
The result is an 
interacting system in which turbulence amplifies the magnetic field  
which, in turn, increases the level of turbulence. Numerical simulations
have shown that such a mechanism can produce a self-sustaining dynamo in
either cyclic or chaotic forms. The form of the dynamo depends on such
physical characteristics as the strength and configuration
of the background magnetic field, the density structure of the disk and
the conditions at the edge of the disk (see, e.g.
Torkelsson \& Brandenburg 1994; Brandenburg et al. 1995; Hawley, Gammie \& 
Balbus 1996). Of particular interest to our study are numerical 
simulations of Brandenburg et al. (1996), which predict a
cyclic dynamo with a period of $\sim$30 times the Keplerian rotation period.
If this simulation is applicable to $\gamma$\,Cas, then the observed 70
day cycle implies that the dynamo is operating in the inner disk,
at a radius of about 2.5\,R$_{\star}$ (assuming M$_{\star}$=17 M$_{\odot}$ and 
R$_{\star}$ = 7\,R$_{\odot}$). This is just slightly outside the Keplerian 
co-rotation radius, R$_K$ = 1.7\,R$_{\star}$ and is near the location where 
the disk density seems to be highest (Berio et al, 1999). 

We suggest that the observed optical variations in $\gamma$\,Cas are
caused by the turbulence generated by the disk dynamo and the effect of that 
turbulence on the density structure of the disk. To understand this process, 
we note that the primary source of energy within the disk, including that
which powers the optical emissions, is the radiation 
field of the star. Since the disk is optically thin in the continuum (e.g.,
Bittar \& Stee 2001), the amount of radiation which is absorbed and re-emitted
is dependent on the local emission measure, $\int n_e^2 dV$. Thus, changes in 
the density will be reflected as variations in disk brightness. Since 
the disk is 
moderately stable over a timescale of years, we stipulate that in the absence
of a dynamo there exists a dynamic equilibrium involving the wind, magnetic 
fields, turbulence, viscosity, etc. which will determine the density 
structure. As discussed extensively by Balbus \& Hawley 
(1998), the introduction of magnetic turbulence into the disk will result
in the outward transfer of angular momentum.
This causes a net inward drift of material and a consequent increase in the 
density of the inner disk, which then increases in brightness.
As the magnetic fields decrease later in the dynamo cycle, the turbulence will 
also decrease, so the disk will evolve back to its original density and 
brightness.

The energy source for dynamos operating in stellar {\em accretion disks} is
ultimately the gravitational energy of the accreted material. In contrast,
we propose that the dynamo within the {\em decretion} disk around
$\gamma$ Cas operates just 
above the Keplerian co-rotation radius, R$_K$ and is supplied with energy by 
magnetic fields which connect the disk to the stellar surface. To see why 
this is necessary, consider that material inside 
R$_K$ must rotate faster than the stellar surface in order to maintain its 
distance from the star. {\em In the absence of other processes}, 
we expect that magnetic fields tied to the stellar surface will 
produce a drag on this material and cause it to spiral onto the star, 
resulting in an inner edge to the disk which is at or near R$_K$. 
Above R$_K$, these same fields will impart energy and angular momentum 
to the gas from the star's rotational reservoir. This action prevents the 
matter within this region from being lost during the maximum of the dynamo
cycle. In this context we note that Berio et al. (1999) find evidence
for a density maximum at a radius of about 2.5R$_{\star}$. Further, 
Smith \& Robinson (1999) report on drifting spectral features near the Si IV
resonance lines which can be interpreted as corotating clouds near 
2.5R$_{\star}$. This provides evidence for the existence of surface connected
magnetic fields at these heights. More evidence comes from Waters et al.(2000),
who found that the hydrogen Pfund, Humphreys, Hansen-Strong, lower level 8 
and lower level 9 emission line series in the infrared spectrum of 
$\gamma$\,Cas all increase in width with decreasing line strength, as expected
in a disk where the rotational velocity decreases with height. The maximum
width of the lines in these series is in the range $500-650$ km~s$^{-1}$.
From the stellar parameters we have adopted, and assuming an 
inclination of 46$^{o}$ of the rotational axis to our line of sight 
(Quirrenbach et al. 1997), we estimate that the Vsin\,i of a Keplerian
disk at the surface of the star is only $\sim$480 km s$^{-1}$. Thus, even 
acknowledging possible errors in our assumed parameters, the large
line widths suggest that the inner part of the disk is
being forced to corotate with the stellar surface, presumably through 
the action of interconnecting magnetic fields. If this is true, then adopting
a Vsin\,i of 230-310 km s$^{-1}$ for the star (e.g., Slettebak, 1982) implies
that the forced corotation extends to a height of about 1.6-2.8 R$_{\star}$, 
i.e. near or just above R$_K$.

\section{Conclusions}

In this paper we have described the results of  monitoring 
$\gamma$\,Cas at both optical and X-ray wavelengths. The graduated spacings 
in the {\it RXTE/PCA} observations
have succeeded in defining a timescale for long-term X-ray variations whose
existence was merely indicated from previous random shorter-term monitorings.
The sequence of rotation-averaged X-ray fluxes conflicts with both the period
and amplitude of the modulation expected from conventional Bondi-Hoyle 
accretion onto the newly discovered binary companion.
Since this is the same theory used to predict the X-ray flux level from 
$\gamma$\,Cas in binary models (e.g., Kubo et al. 1998), this disagreement 
poses manifest problems for the binary accretion hypothesis. 

The individual X-ray light curves also show ``features" which indicate 
the presence of evolving active centers. Comparison of light curves
shows that these structures evolve in time and probably migrate across 
the stellar surface. Thus, using the X-ray signatures as markers is not a 
reliable method for determining the star's true rotation period. It is
unclear whether this conclusion affects our earlier determination of the
period, mainly from UV markers.

Probably the most surprising result of the study was the discovery of cyclic
variations in both the X-rays and optical fluxes. A study of these variation 
shows that:

\begin{itemize}

\item the X-rays and optical fluxes appear to have a strong positive 
correlation, 

\item the cycle length changes with time, with observed values ranging from 
55 to 93 days,

\item the amplitude of the optical variations is greater in the $V$ than 
 the $B$ band. 

\end{itemize}

In previous papers of this series we have suggested that the X-rays from 
$\gamma$\,Cas arise from plasma heated by magnetic instabilities and stresses. 
The results of the current work indirectly confirms this hypothesis by 
providing evidence against the binary star hypothesis, which appears to be 
the only other viable alternative capable of explaining observed X-ray fluxes 
and temperatures. 

   The short timescales of X-ray flares (often as short as a few seconds, 
SRC98) imply that this emission is formed in a relatively high density 
region, probably the upper photosphere. The location of the basal X-ray 
component is somewhat more uncertain, though there are indications that 
it is also produced near the star. However, the properties of the optical 
variability, particularly the increase in $(B-V)$ color with increasing flux 
and the lack of correlated changes in the UV flux, suggest that these are
generated within the Be disk and not from a change in photospheric flux.
The optical fluctuations contain far too much energy to be caused by the 
reprocessing of X-rays. In view of this fact we suggest that the
correlated X-ray and optical variations are both driven by a cyclical
process, which we conjecture is due to a magnetic dynamo, and probably
operates within the disk, 
Such a dynamo would provide a time-modulated mean magnetic 
field in the disk that drives X-ray activity and is also associated with the 
production of magnetic turbulence through the Balbus-Hawley instability. This
turbulence influences the transport of angular momentum within the disk and 
will influence the density structure, and therefore the brightness, of the 
disk. Thus, the control of the X-ray and optical variations arises from two
separate but related properties of the dynamo.
The study of disk dynamos is still in its infancy. Most studies to date have
been specific to accretion disks rather than the decretion disk of Be stars 
such as $\gamma$\,Cas. However, at least some {\it ab initio} calculations, 
while specific to special conditions, do show that cyclic disk dynamos 
are possible and can vary on a timescale comparable to that reported here. 
Thus, the idea of a disk dynamo on $\gamma$\,Cas should be pursued.

 At this point it is worthwhile summarizing the ``big picture"  
emerging from the series of papers starting with SRC98 concerning the 
production of anomalous X-rays 
in $\gamma$\,Cas. In this picture a highly complex magnetic topology 
exists on the surface of the star. These fields evolve rapidly
and may also migrate across the stellar surface. During times of low X-ray flux
the fields appear to be concentrated into 2 or 3 complexes. However, when the
X-rays are near maximum the fields are more evenly spread across the disk, 
and the X-ray flux shows a smaller rotational modulation. A key assumption 
in our picture is that the star's magnetic field becomes entrained in the 
inner part of the ionized, circumstellar 
disk. This interaction has two consequences. First, the stellar 
magnetic field causes the production of turbulence 
within the disk through the Balbus-Hawley instability. This turbulence 
amplifies
and modulates the stellar ``seed" field through a disk dynamo. The turbulence
also affects the density structure of the inner disk, causing the disk 
brightness to change. Second, the difference in 
angular rates of rotation between the star and disk results in the stressing 
and shearing of the magnetic lines of force. This causes the ejection of high 
velocity plasmoids (similar to solar ``coronal mass ejections," Smith \& 
Robinson 1999) as well as the generation of high-energy particle beams, 
some of which are directed toward particular regions on the star. 
The impact of these beams on the surface results in 
explosive heating of the ambient plasma to a temperature near 10$^8$\,K, 
resulting in the X-ray shots. The
rapid expansion and entrapment of this plasma along overlying magnetic field 
lines results in the longer lived  basal X-ray emission, which usually 
accounts for most of the X-ray flux we observe. 
Clouds of translucent material also form in these same magnetic complexes and 
are responsible for periodic absorptions in the UVC flux. The ionization state 
of particles in various parcels of 
the star's radiative wind is also modulated as they ``see"
these X-ray generating centers (Cranmer, Smith, \& Robinson 2000).
Altogether, the disk serves as a conduit 
in which the star's rotational energy is converted to a 
time-dependent magnetic field and turbulence. The field is dissipated, 
in part, by the ejection of plasmoids and particle beams. Note that 
the optical variations are initiated by the motions set up by the dynamo 
but are not strictly powered by it.

  What are the implications of this picture for this and other stars 
observed over a period of time? 
We do not really know whether the production of X-rays through this
complicated mechanism scales with the development of the disk or is 
initiated above some threshold. However, we can speculate that the X-rays 
of $\gamma$\,Cas were not present ca. 1937 when the disk was very weak 
or non-existent. At the time the X-rays were first 
discovered in late 1975 (Jernigan 1976, Mason et al. 1976), the current disk 
phase of $\gamma$\,Cas was already well underway and similar to its present 
state of development (Horaguchi et al. 1994). Ultimately, and on an unknown 
timescale, the current disk {\em will} dissipate, and so from our picture
one anticipates that the X-ray emission will fade as well. 
Because the engine we outline is complicated 
and because some X-ray properties of $\gamma$\,Cas itself are 
unpredictably (so far) time-dependent, it is unlikely that a $\gamma$\,Cas 
analog would show identical characteristics to $\gamma$\,Cas itself during 
the times we have observed it. For example, we would be mildly surprised if 
an analog should exhibit a 70-day modulation, and its X-ray light curve may 
or may not exhibit a clear rotational modulation at any given time. 
However, the production of hard X-rays and shots is expected to be a 
distinguishing characteristic. In this context, we suggest that the X-ray 
variable source HD\,110432 (B0.5 IIIe) might be a suitable candidate for 
observation. 
Reporting on BeppoSAX/MECS observations, Torrejon \& Orr (2001) have
found that this source shows variable X-rays on timescales of 
4 hours and (perhaps) a few minutes or less, a L$_{x}$ = 7$\times$10$^{32}$ 
erg~s$^{-1}$ (2-10 keV) and, importantly, a hard X-ray spectrum consistent 
with a temperature of 10.55 keV. From its strong He\,I $\lambda$4471
emission and energy distribution in the UV and optical, it is also clear
that the star has a strongly developed disk compared to most
other Be stars (Zorec, Ballereau, \& Chauville 2000, Moujtahid \& Zorec 2000). 
Moreover, it appears in the same region of the H-R Diagram as $\gamma$\,Cas,
is a rapid rotator (Codina et al. 1996), and may exhibit 
variability on a rotation timescale in the optical region (Barrera, Mennickent 
\& Vogt 1991). In these ways the two stars appear to be nearly twins. 
A confirmation of even one more star which shows X-ray characteristics similar 
to the so-far unique case of $\gamma$\,Cas would be critical to 
establishing the range of stellar and disk characteristics responsible for 
this rare X-ray phenomenology and to testing the plausibility of 
competing models. \\ \\

\noindent
It is our pleasure to thank Karen Bjorkman and Anatoly
Miroshnichenko for putting their Ritter Observatory H$\alpha$ observations
of $\gamma$\,Cas at our disposal in advance of their formal publication.
We also appreciate an H$\alpha$ spectrum obtained for us in 2000 by G. Peters. 
We gratefully acknowledge helpful suggestions by Petr Harmanec and informative 
theoretical discussions on the disk dynamo problem by S. Owocki, J. Stone, 
and E. Vishniac. We are also indebted to Philippe Stee for repeating a 
Bittar-Stee model to predict a disk contribution to integrated UV flux. GWH 
acknowledges support from NASA grants NCC5-511 and NGC5-96 as well as NSF 
grant HRD-9706268. RDR and MAS acknowledge support from NASA grant 
NAG5-11705.

\clearpage

\figcaption[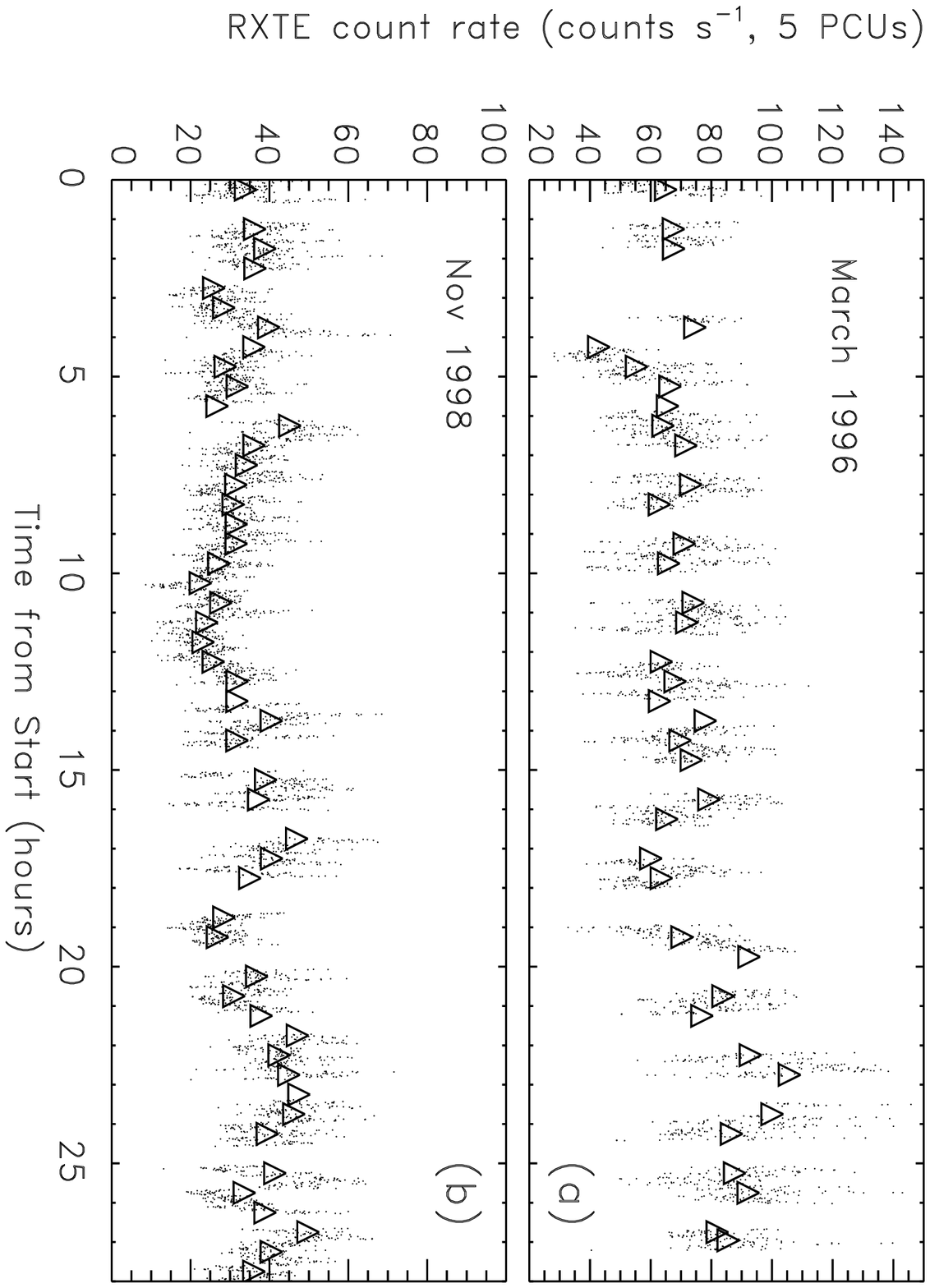]{Summary of {\it RXTE} observations taken in (a)
 March 1996 and (b) November 1998. Points represent 16~s averages, while 
triangles are 30 minute averages. The uncertainties in the data can be 
represented
by Poisson statistics. Thus errors in the position of the triangles
are much smaller than the symbols.  Each plot represents a single 27-hour 
rotation cycle.
\label{lc9698}}

\figcaption[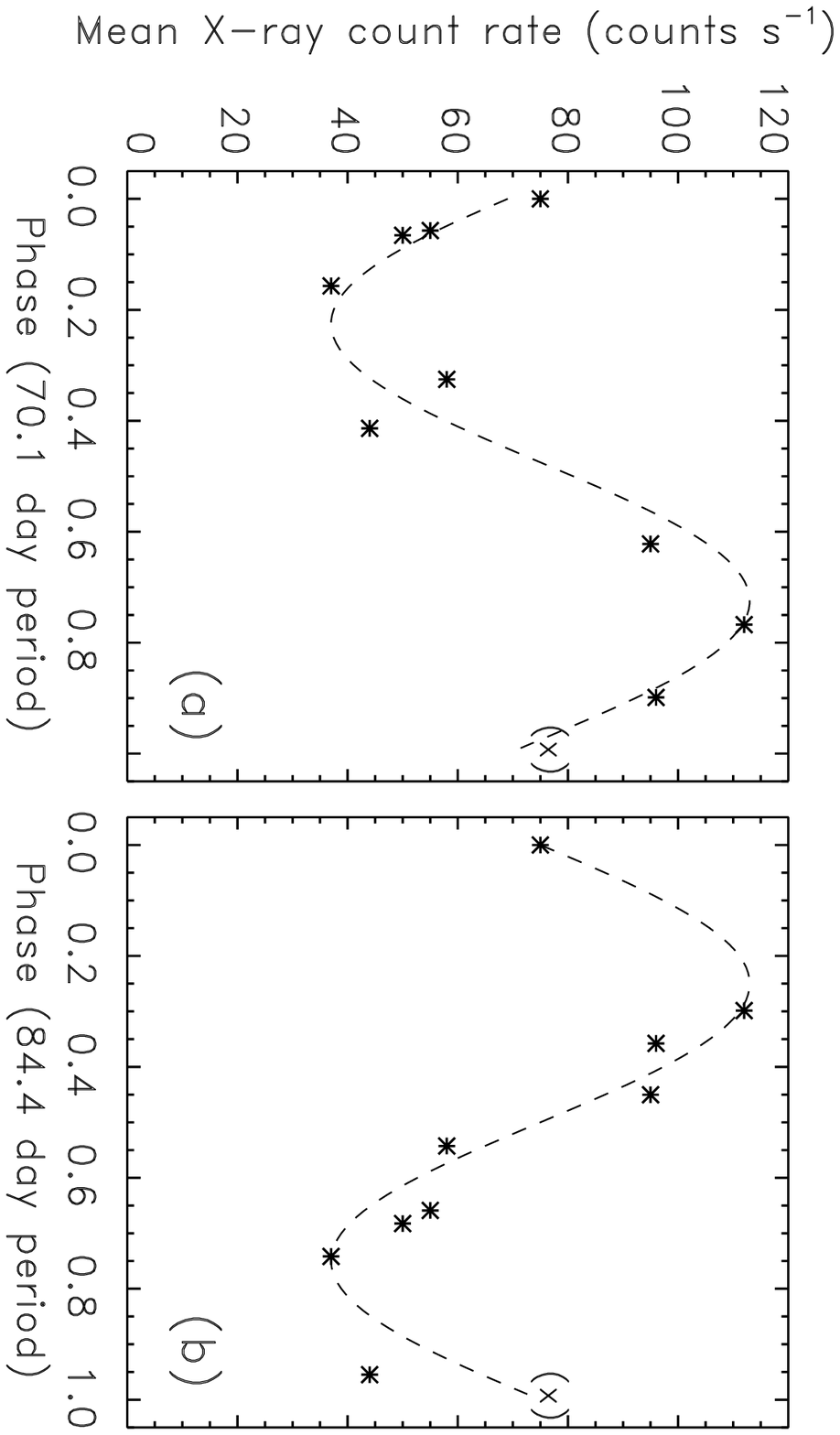]{Median X-ray count rate for the 
9 observed times sequences (asterisk) as a function of cycle phase, 
assuming (a) a 70.1 day cycle period and (b) an 84.4 day period. The point
at phase 0 has been replotted at phase 1 for reference. A sine wave (dashed
line) has been included for reference. 
\label{ltxvar}}

\figcaption[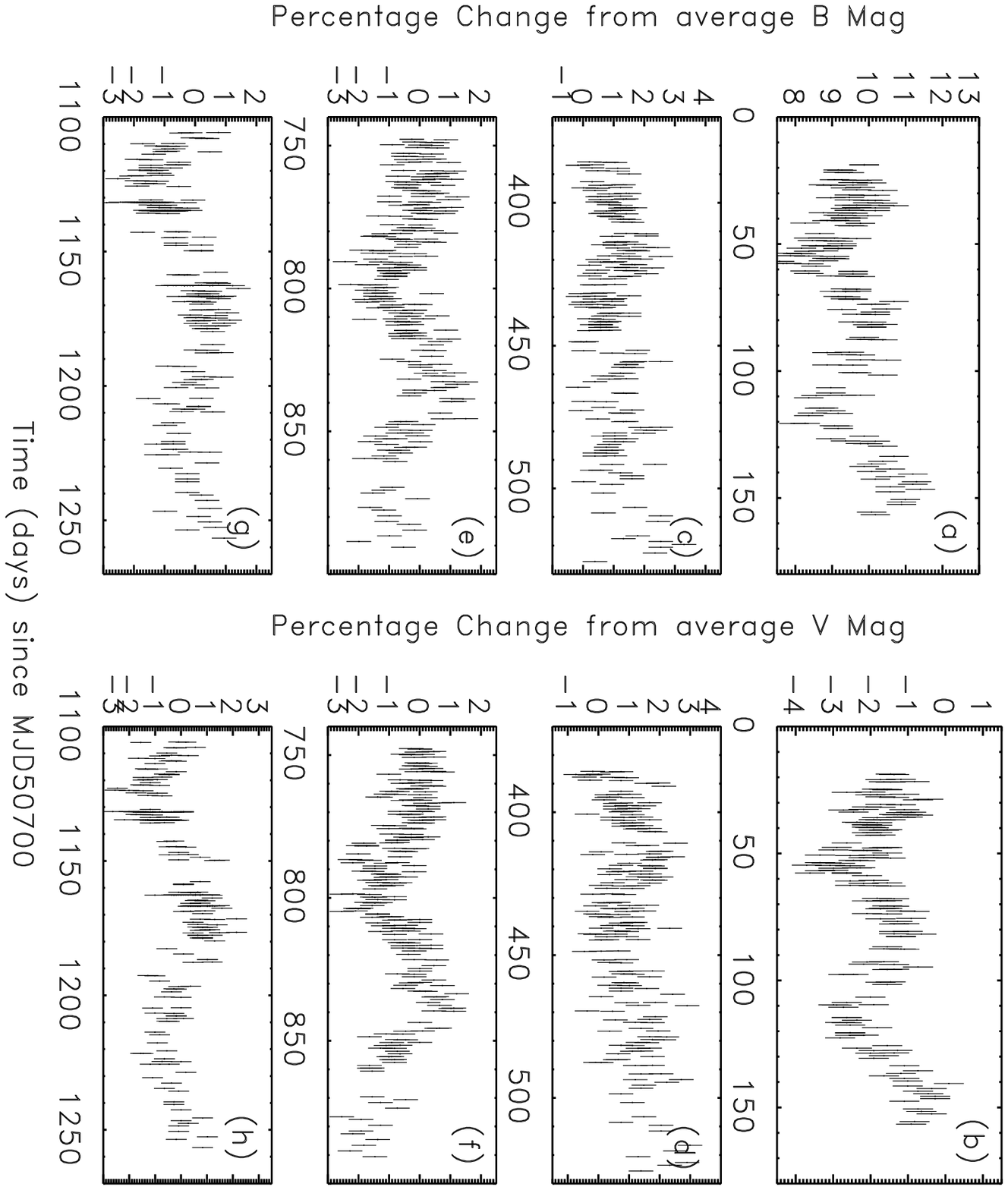]{Summary of the optical $B$ and $V$ band
observations for each of the observing seasons. $B$ and $V$ magnitudes 
have been 
converted to percentage deviation from the average magnitude derived over
the entire data set. 
\label{optsum}}

\figcaption[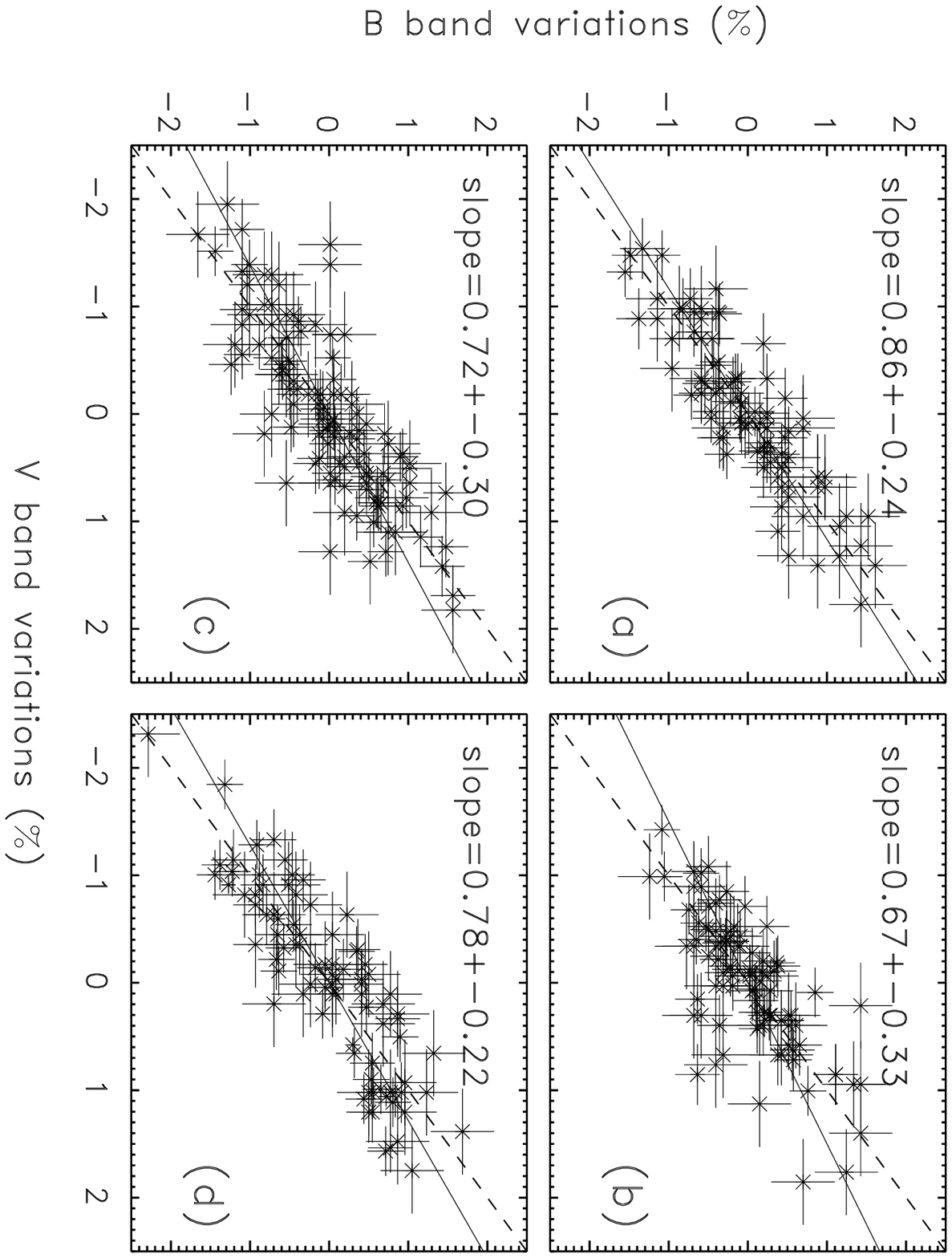]{Comparing $B$ and $V$ band variations 
during each 
of the 4 APT observing seasons: (a) 1997-1998, (b) 1998-1999, (c) 1999-2000,
(d) 2000-2001. Points represent intensities averaged over 1-day intervals 
and expressed as a percentage deviation from the mean intensity for that 
observing season. The solid line is a linear least squares fit to the
data ("slope" indicates the proportionality factor between $\Delta$B and
$\Delta$V), while the dashed line represents the case where $\Delta$B 
equals $\Delta$V (no color changes). 
\label{bvcomp}}

\figcaption[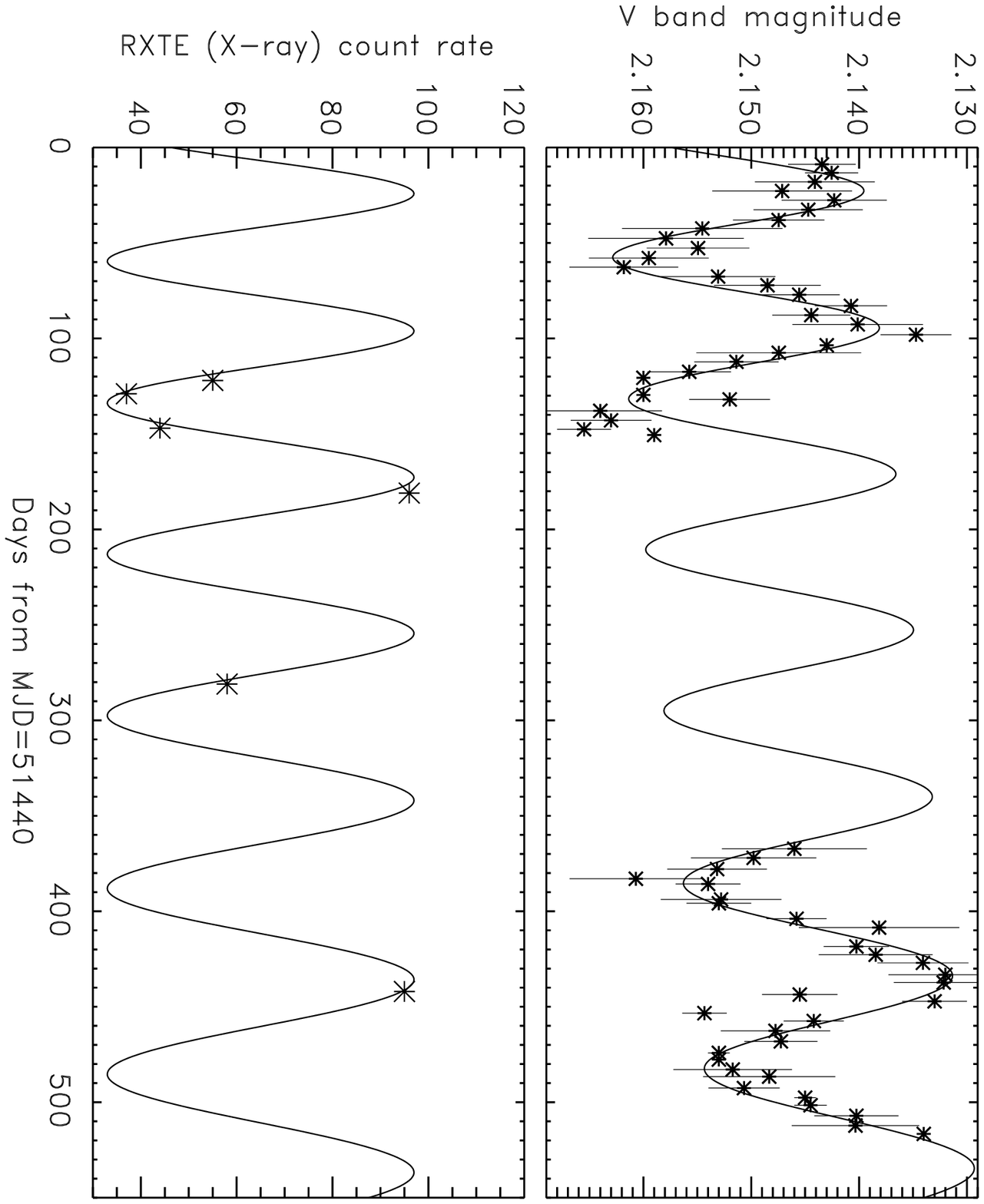]{(top) 5 day averages of $V$ band magnitude
over the 1999-2000 and 2000-2001 seasons. Error bars represent the rms 
variation of the data points. The solid line is an empirical fit to a sin 
wave with linearly increasing period and phase, as explained in the text.
(bottom) Median {\it RXTE} count rates for the 6 time sequences 
obtained in 2000.
The solid line is the sin wave model used in the optical fit, which has been
adjusted in amplitude to fit the X-ray data. 
\label{optx}}

\figcaption[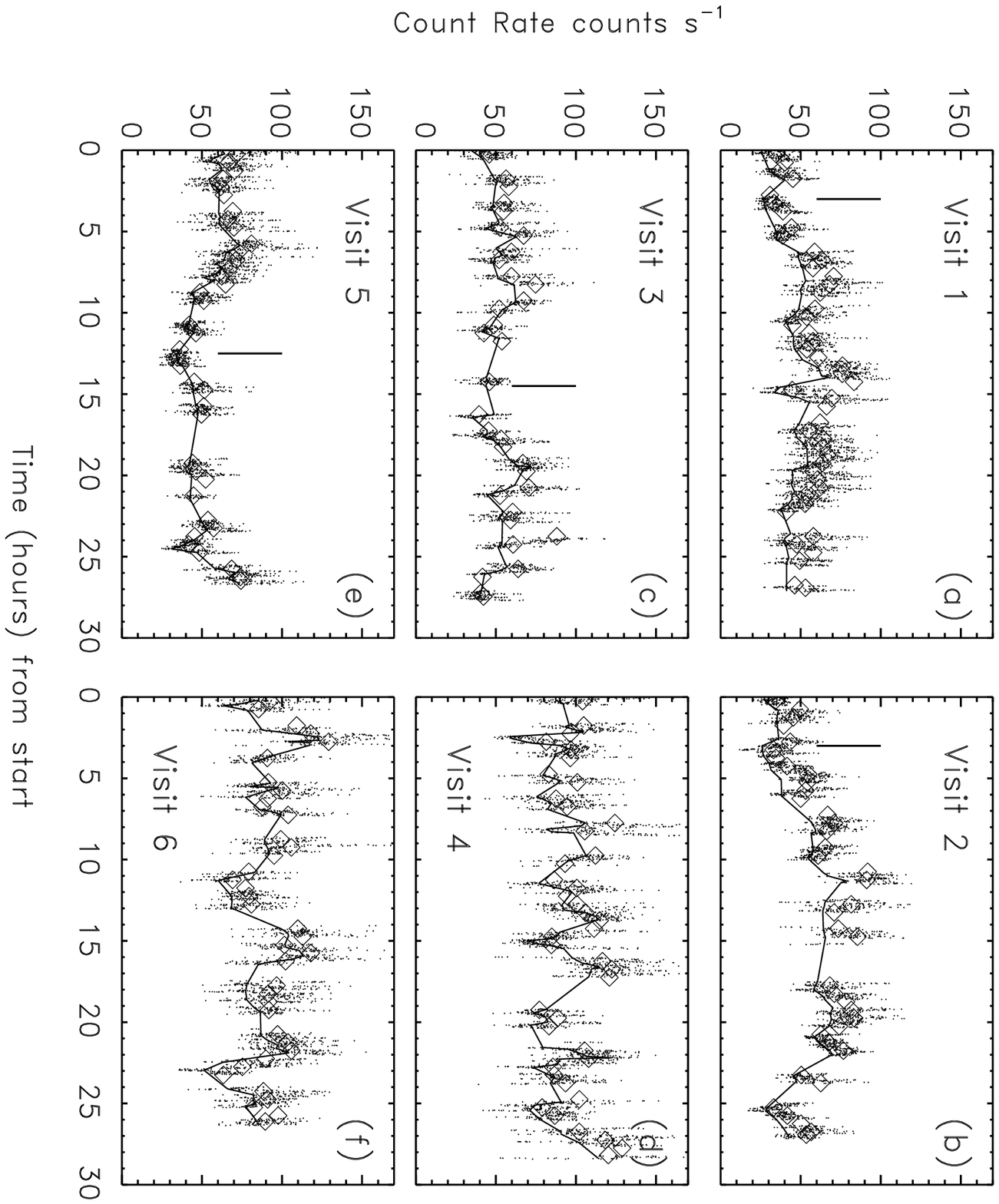]{Summary of the 6 {\it RXTE} time sequences
obtained in 2000. Points represent 16~s averages, diamonds are 30 minute 
averages and the solid line represents estimates of the basal flux level.
A vertical line in visits 1, 2, 3 and 5 shows the location of a flux
minimum which was used to search for a rotation period of the star (see
Figure~\ref{mincomp}).
\label{xsum}}

\figcaption[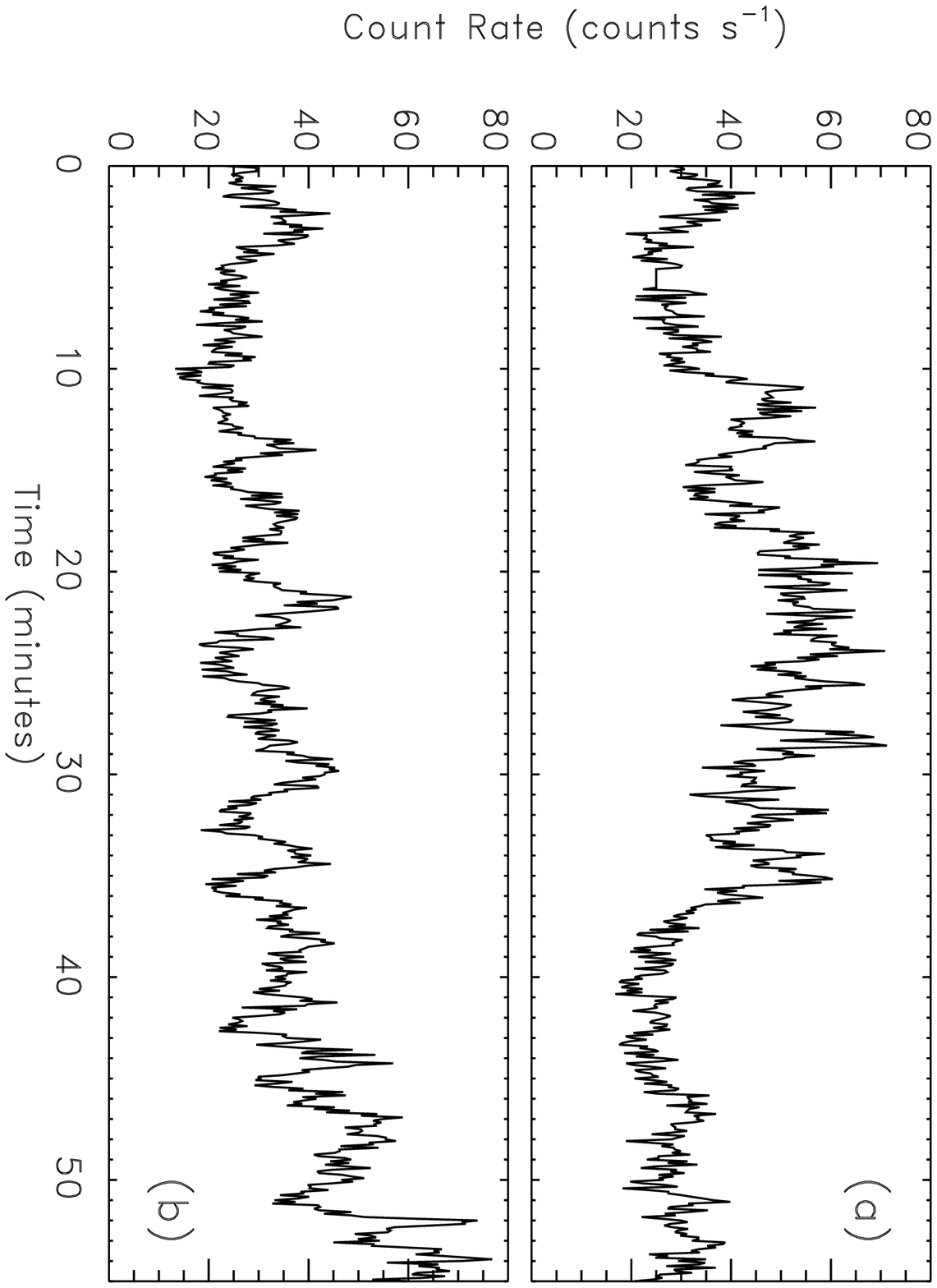]{Detail of the {\it RXTE} time series at two 
representative times which shows both the short term transients (shots) and 
the longer term variations in the underlying basal emission.
\label{xshort}}

\figcaption[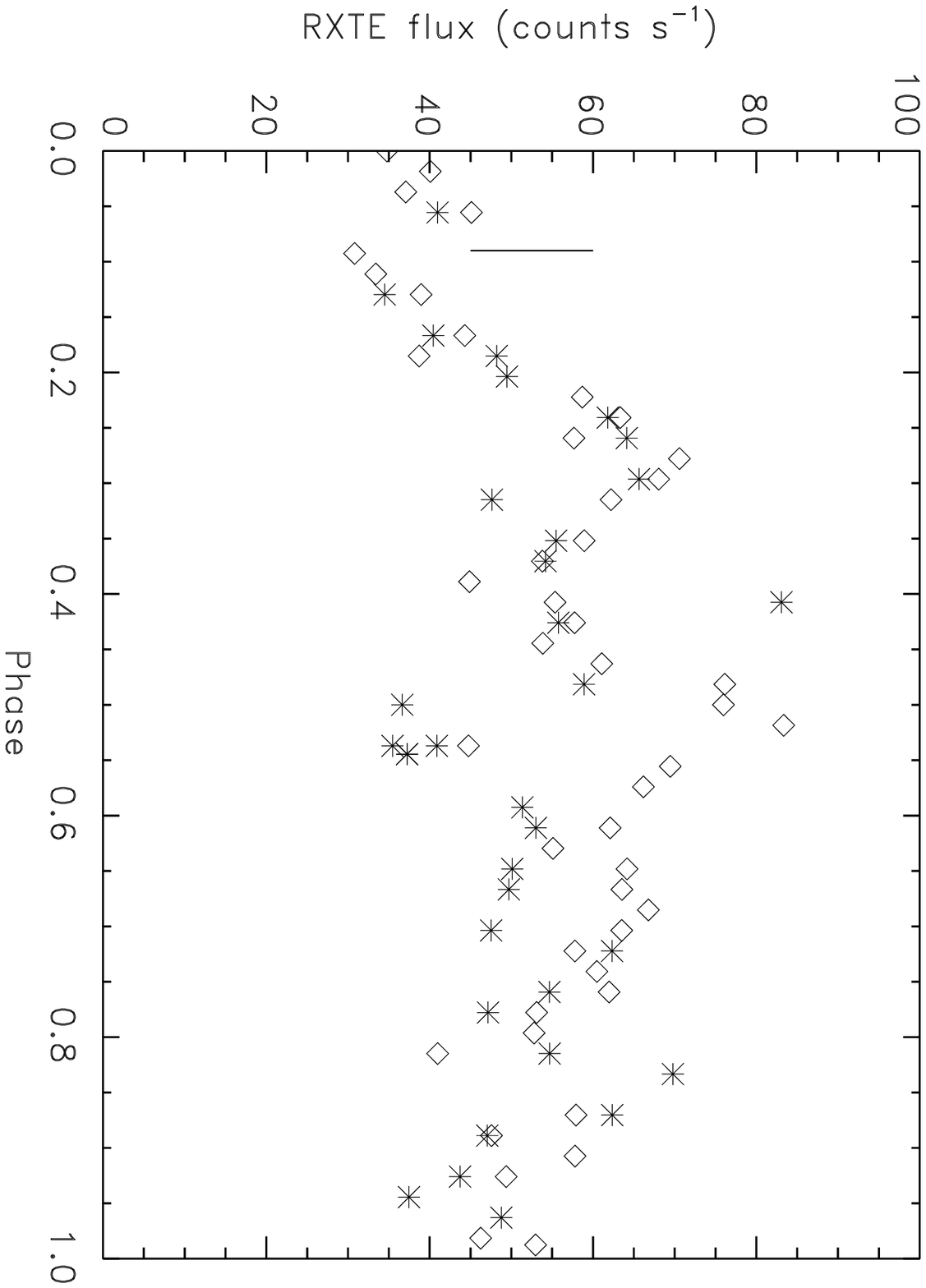]{Comparing the 30 minute average 
X-ray fluxes obtained during visit 1 (diamonds) with average fluxes obtained
during visit 3 (asterisks). The points are plotted as a function of rotation
phase assuming a 27 hour period. The fluxes from visit 3 have been shifted
in phase to align the flux minimum feature seen at phase 0.1.
\label{mincomp}}

\figcaption[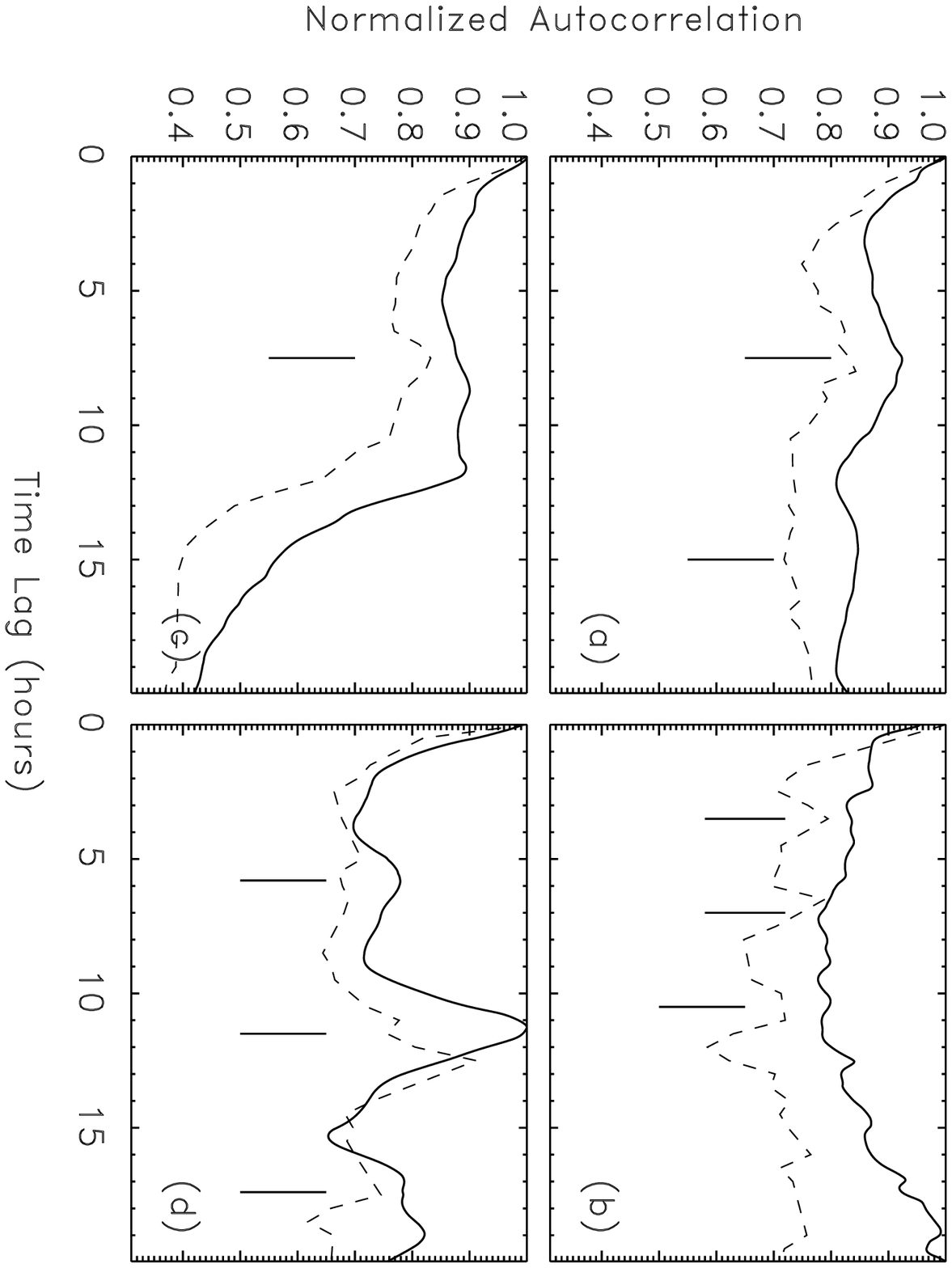]{Autocorrelation functions for the reciprocal
X-ray flux obtained from various time sequences: (a) Nov 1998, (b) visit 4 
(2000 Mar 17), (c) Visit 5 (2000 June 25), (d) visit 6 (2000 Dec 03). The 
solid line is a calculation
obtained from the basal flux measurements, while the dashed line is the 
same calculation using 30 minute average fluxes (which include the effects of 
shots). Vertical lines indicate equally spaced peaks in the correlation which
point to periodic behavior. The indicated periods are (a) 7.5 hours, 
(b) 3.5 hours, (c) $\sim$7 hours and (d) 5.8 hours. 
\label{invccor}}

\figcaption[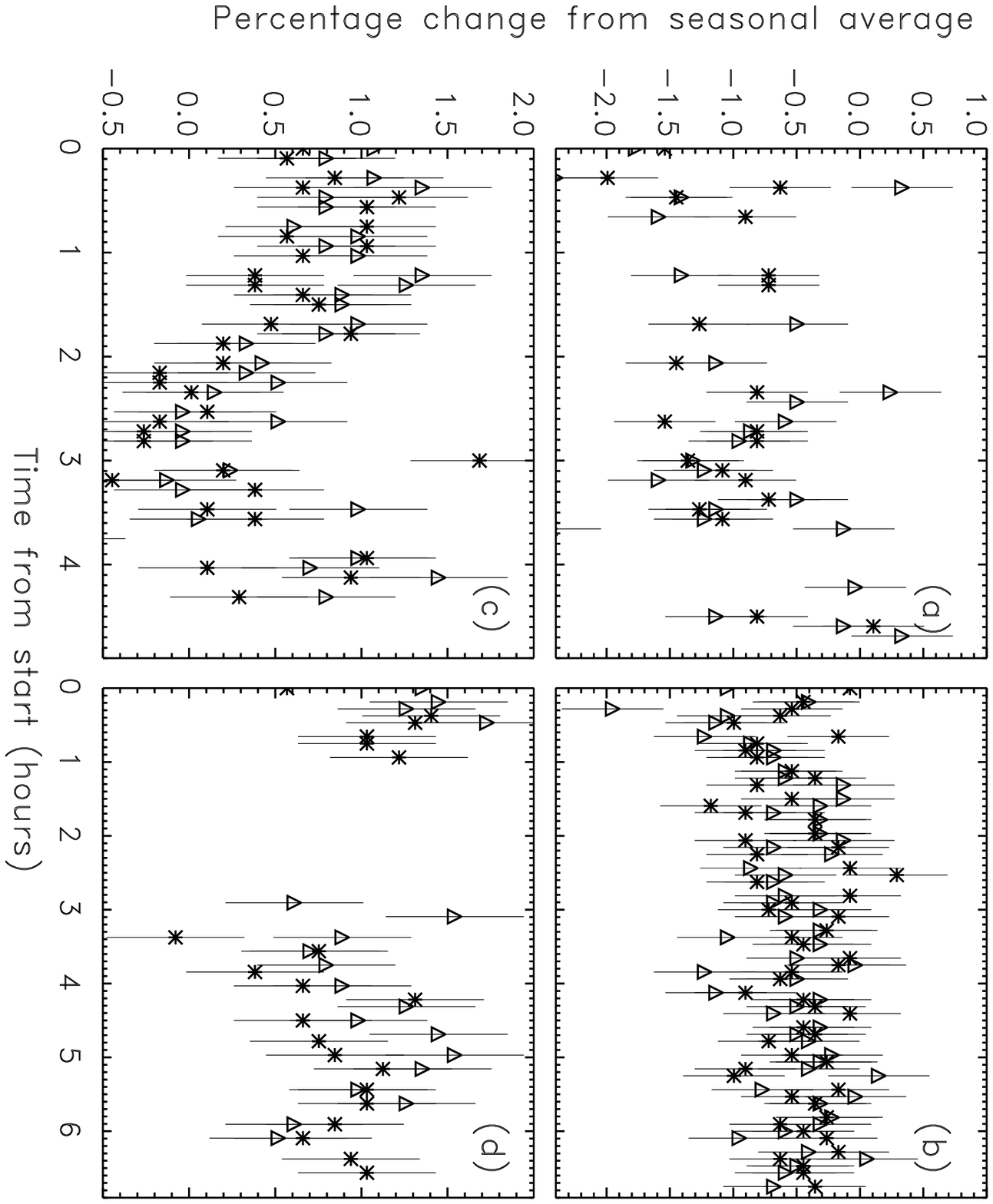]{Short term optical variations
obtained during the 2000-2001 observing season. Fluxes are represented as 
the percentage change from the mean magnitude seen during that observing 
season. Diamonds represent $V$ band observations, while asterisks 
mark $B$ band data. Observing dates are (a) 2000 Oct 14, (b) 2000 Oct 17, 
(c) 2000 Nov 14, (d) 2000 Nov 18.  
\label{shortopt}}

\figcaption[fig11.ps]{Mean {\it RXTE} fluxes folded to the Harmanec
et al. period of 203.59 days. Their ephemerides are applied to Bondi-Hoyle
accretion theory (with scaling factor adjusted to mean of observed fluxes)
to produce the dashed curve, which clearly does not fit the data.
\label{bincomp}}

\clearpage
\begin{center}
{\bf TABLE 1} \\
{\bf Summary of {\it RXTE/PCA} Observations} \\ 
\begin{tabular}{ccccccc} \hline
Visit & Date & Date & Duration & Count Rate & Position of \\
  & (HJD - 2,450,000) & (2000) & (hours)    
  & (counts s$^{-1}$) & minimum (hr)$^a$ \\ \hline \\
1 & 1562 & 18 Jan &27.16 & 55 & 0 \\
2 & 1569 & 25 Jan & 27.17 & 37 & 158.0 \\
3 & 1587 & 12 Feb & 27.7 & 44 & 589.0 \\
4 & 1621 & 17 March & 28.4 & 96 & -- \\
5 & 1721 & 25 June & 26.7 & 58 & 3808.9 \\
6 & 1882 & 03 Dec & 26.4 & 95 & -- \\
\hline
\end{tabular}
\end{center}
$^a$ Time from the minimum measured in visit 1 \\

\clearpage
\begin{center}
{\bf TABLE 2} \\
{\bf Summary of Optical Observations$^a$} \\ 
\begin{tabular}{cccccc} \hline
Photometric &  Date Range  &   Nobs  &   Period  &    Ave   &  Peak-to-Peak \\ 
   Band  &   (HJD-2,450,000) &  &     (days)  &   Mag  &     Amplitude \\ 
     &                   &  &             &        &      (mag) \\ \hline \\
 $B$ &  0718 - 0856  &   177 &  60.8 $\pm$ 1.3 & 1.991  & 0.0142 $\pm$ 
0.0012  \\
 $V$ &  0718 - 0856  &   183 &  61.7 $\pm$.3 & 2.163  & 0.0156 $\pm$ 0.0012  
\\ \\
 $B$ &  1086 - 1225  &   207 &  52.8 $\pm$ 0.8 & 2.078  & 0.0086 $\pm$ 0.0012 
 \\ 
 $V$ &  1085 - 1225  &   211 &  55.0 $\pm$ 0.9 & 2.132  & 0.0089 $\pm$ 0.0014
  \\ \\
 $B$ &  1447 - 1590  &   247 &  72.2 $\pm$ 1.6 & 2.093  & 0.0154 $\pm$ 0.0012 
 \\ 
 $V$ &  1447 - 1590  &   243 &  77.1 $\pm$ 1.8 & 2.150  & 0.0174 $\pm$ 0.0011 
 \\ 
\\
 $B$ &  1805 - 1956  &   290 &  93.2 $\pm$ 3.0 & 2.094  & 0.0172 $\pm$ 0.0010 
 \\ 
 $V$ &  1805 - 1956  &   290 &  93.6 $\pm$ 2.8 & 2.146  & 0.0200 $\pm$ 0.0009 
 \\
\hline
\end{tabular}
\end{center}
$^a$ The individual Johnson BV photometric observations are available at \\
http://schwab.tsuniv.edu/t3/gammacas/gammacas.html

\end{document}